\documentclass[%
 reprint,
 amsmath,amssymb,
 aps,
]{revtex4-2}

\usepackage{graphicx}
\usepackage{dcolumn}
\usepackage{bm}
\usepackage{physics}
\usepackage{multirow}

\usepackage{dsfont}
\usepackage[colorlinks=true,linkcolor=blue,citecolor=blue,urlcolor=blue]{hyperref}

\usepackage{algorithm}
\usepackage{algpseudocode}

\begin{document}


\title{Efficient fidelity evaluation in few-jump limit of the Monte Carlo Quantum Jump method}
\title{Low Variance Quantum Jump Algorithm in the Few Jump Limit}
\title{Deterministic Quantum Jump (DQJ) Method  for Weakly Dissipative Systems}


\author{Marcus Meschede,$^1$}
\email{marcus.meschede-2@uni-hamburg.de}

\author{Ludwig Mathey$^{1,2}$}

\affiliation{
$^1$Center for Optical Quantum Technologies and Institute for Quantum Physics, University of Hamburg,
22761 Hamburg, Germany\\
$^2$The Hamburg Center for Ultrafast Imaging, Luruper Chaussee 149, 22761 Hamburg, Germany\\
}

\date{\today}

\begin{abstract}

Physical quantum systems are generically coupled to an environment, resulting in open system dynamics. 
A typical approach to simulating this dynamics is to propagate the density matrix of the system via the Lindblad master equation. This approach is numerically challenging due to the size of the density matrix, which has led to the development of quantum jump methods, which unravel the density matrix into an ensemble of state vectors. These methods utilize a stochastic sampling of the quantum jump times, which becomes inefficent for weakly dissipative dynamics, in which jumps are rare events. Here, we propose the deterministic quantum jump (DQJ) method, which we show to outperform standard quantum jump methods in the weakly dissipative regime, by removing the error of stochastic sampling. We describe the methodology at the single-jump and two-jump level, reconstructing the density matrix at the corresponding level. We demonstrate the performance of the method for two examples, the dissipative transverse-field Ising model, and the dissipative Kerr oscillator. Given that quantum technologies such as quantum computing have weakly dissipative quantum dynamics as their central focus, we propose this method to be utilized in that context, for exploring and understanding quantum technology platforms.

\end{abstract}

\maketitle



\section{Introduction}
\label{sec:Introduction}

Open quantum systems describe quantum systems coupled to an external environment. They arise in a vast range of physical systems, such as atomic, photonic or solid-state systems \cite{breuerTheoryOpenQuantum2007, zollerQuantumNoiseQuantum1997, weissQuantumDissipativeSystems2012}. 

In the regime of weak system-bath coupling, and for sufficiently Markovian bath dynamics,  a widely used approach is the description via the Lindblad master equation \cite{goriniCompletelyPositiveDynamical1976, lindbladGeneratorsQuantumDynamical1976}. This equation propagates the density matrix of the system in time, under Hamiltonian of the system and dissipative terms composed of Lindblad operators.

However, the numerical integration of this equation becomes computationally intractable with increasing size of the Hilbert space because of the size of the density matrix. Quantum jump methods tackle this problem by unraveling the evolution of the density matrix in an ensemble of pure state trajectories. In its original formulation, the trajectories are determined by propagating the state with a non-Hermitian effective Hamiltonian and applying jump operators at stochastically sampled jump times. In the following we refer to these baseline formulations as the standard quantum jump (SQJ) methods.  \cite{cohen-tannoudjiSingleAtomLaserSpectroscopy1986, dalibardWavefunctionApproachDissipative1992, dumMonteCarloSimulation1992, dumMonteCarloSimulation1992a, carmichaelOpenSystemsApproach1993, molmerMonteCarloWavefunctions1996} . 

Beyond the core jump method, methodological improvements have broadened the scope of the quantum jump method
\cite{steinbachHighorderUnravelingMaster1995, piiloNonMarkovianQuantumJumps2008, kornyikMonteCarloWavefunction2019, abdelhafezGradientbasedOptimalControl2019, macieszczakQuantumJumpMonte2021,radaelliGillespieAlgorithmQuantum2024} and the jump count statistics have been shown to connect well to physical quantities \cite{garrahanThermodynamicsQuantumJump2010}

The exploration and development of weakly dissipative system is essential for quantum technologies. For instance, quantum computing platforms such as trapped ions \cite{haffnerQuantumComputingTrapped2008, foss-feigProgressTrappedIonQuantum2025}, cQED setups \cite{devoretSuperconductingCircuitsQuantum2013, jiangAdvancementsSuperconductingQuantum2025} and, spin-qubit setups \cite{onizhukColloquiumDecoherenceSolidstate2025} operate in the regime of very weak dissipation, necessarily. To simulate the dynamics in this regime numerically, aids the development of pulse engineering \cite{khanejaOptimalControlCoupled2005, kochQuantumOptimalControl2022}, variational quantum algorithms \cite{peruzzoVariationalEigenvalueSolver2014, cerezoVariationalQuantumAlgorithms2021} or error-mitigation \cite{temmeErrorMitigationShortDepth2017, caiQuantumErrorMitigation2023}. However, in this regime the standard quantum jump methods face the challenge of quantum jumps occurring as rare events, making a stochastic sampling inefficient.

To address this challenge, we introduce the deterministic quantum jump method (DQJ), that outperforms existing quantum jump methods in the weak dissipation regime of the Lindblad master equation. DQJ eliminates the stochastic error of SQJ, by placing the quantum jumps deterministically on a suitably chosen jump time grid. We show both analytically and numerically, that our method scales better in the number of computed jump trajectories than stochastic jump methods in reconstructing the approximate density matrix. Our method is only limited by the magnitude of the neglected jump order in the DQJ method, which constitutes a well-controlled and quantifiable error. For a physical system with intrinsic dissipation rates $\gamma_j$, the weakly dissipative regime corresponds to $\gamma_j T \ll 1$, for all $j$, where $T$ is the total time of the dynamical evolution. Given that quantum technological platforms fulfill this requirement,our method is ideally suited for the simulation of these systems.

This paper is organized as follows. In Sec.~\ref{sec:Method_a}, we introduce the single jump DQJ method and provide a pseudocode algorithm to show how to implement the method. In Sec.~\ref{sec:Method_b}, the two jump DQJ method is described. The applicable regime for the DQJ method is discussed in  Sec.~\ref{sec:Method_c}. We illustrate our method with (i) the fidelity of a transverse-field Ising model time evolution and (ii) the Fourier spectrum of the Kerr-resonator in Sec.~\ref{sec:Examples}. Finally, we summarize our results and give an outlook for our method in Sec.~\ref{sec:Conclusion}.

\section{DQJ method at the single-jump level}
\label{sec:Method_a}

\begin{figure*}[ht]
    \centering
    \includegraphics[width=\textwidth]{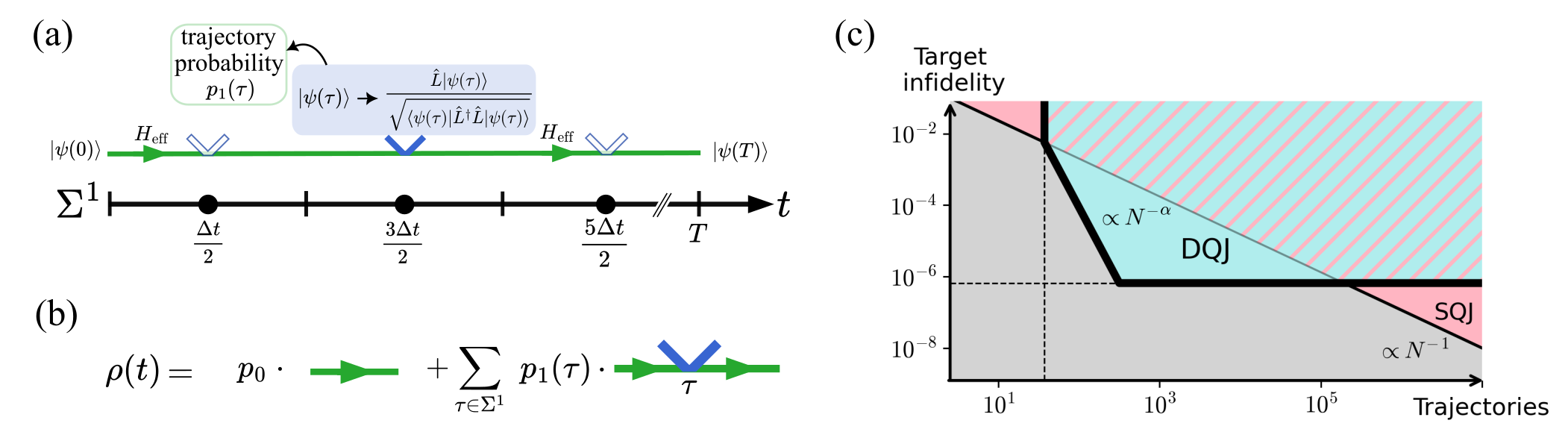}  
    \caption{(a) Schematic generation of a single jump trajectory in the Deterministic Quantum Jump (DQJ) method. The initial state is propagated  to the jump time $\tau$ with the effective Hamiltonian $H_{\text{eff}}$ (green). The jump time is an element of the deterministic and equally spaced jump time grid $\Sigma^1$ defined on the integration domain $t\in[0,T]$. The trajectory probability is recorded and a quantum jump performed (blue). After normalization, the state is  propagated to the final time $T$. (b) DQJ calculates the density matrix evolution by accounting for all jump orders individually. Each jump trajectory with a jump from $\tau \in \Sigma^1$ is weighted by its trajectory probability in the interval $[0,T]$. (c) Regime of the DQJ  and SQJ (red) method as a function of computational trajectories and target infidelity. DQJ outperforms (cyan) or is on par (striped) with the SQJ method (red) if the jump time grid spacing is smaller than the fastest density matrix entry evolution, $\Delta t< 1/f_{\max}$ (vertical dashed line) and higher jump orders are negligible (horizontal dashed line). The power-law exponent of the infidelity is $\alpha = 4/n$ for the $n$ jump DQJ method. SQJ scales much slower in the number of trajectories.}
    \label{fig:Method}
\end{figure*}

Quantum jump methods approximate the time evolution of the density matrix $\rho(t)$ via an ensemble of trajectories of states $\ket{\psi(t)}$. For the density matrix dynamics governed by the Lindblad master equation, the trajectories are formed from initial states $\ket{\psi (0)}$ that evolve with an effective Hamiltonian 

\begin{equation}
    H_\text{eff} = H - \frac{i}{2}\sum_{\hat{L}\in J^1} \hat{L}^\dagger \hat{L} \quad
\label{equ:Heff}
\end{equation}
and which are subject to quantum jumps, in addition to the continuous evolution under $H_{\text{eff}}$. The initial state $\ket{\psi(t)}$ is either the initial state of the physical system, if it is in a pure state, or it is a component of a mixed state, where the components are to be averaged over. Throughout the paper, we will focus on initially pure states. 

The effective Hamiltonian in Eq.~\ref{equ:Heff} contains the Hamiltonian $H$, which describes the unitary time evolution, and a sum over the jump operators $\hat{L}_j$, which are elements of the set of jump operators $J= \{\hat{L}_1, \hat{L}_2, \dots\}$ . We note that the operators $\hat{L}_j$ have units of the inverse-squareroot of time, in contrast to the alternative convention of separating the decay rate from a unitless operator. We introduce the representation $L_j= \sqrt{\gamma_j} \hat{A}_j$, as an explicit form that includes the decay rates $\gamma_j$. In a physical context, the decay rates $\gamma_j$ have a physical origin and meaning, thus justifying the factorization of $\hat{L}_j$ into $\gamma_j$ and $\hat{A}_j$.

In the Standard Quantum Jump (SQJ) method, a quantum jump operator from the set $J$ is applied stochastically with an instantaneous jump rate $n_L(\tau) =  \langle \hat{L}^\dagger \hat{L} (\tau)\rangle$. The ensemble average over these individual trajectories recovers the density matrix evolution. The target infidelity on the approximated density matrix $\rho(t)$ described below, determined through the SQJ method, scales inversely to the number of trajectories N, i.e., $\propto 1/N$ as indicated in Fig.~\ref{fig:Method}(c).

In the weakly dissipative regime of Lindbladian systems we propose the more efficient Deterministic Quantum Jump (DQJ) method which converges to the exact evolution using fewer trajectories. The weakly dissipative regime of the time evolution over a time interval $T$, corresponds to $\int_0^T dt \sum_L \langle L^\dagger L\rangle \ll 1$. Approximately, this corresponds to $\gamma_j T \ll 1$, for all $j$. In this limit, trajectories with few jumps dominate the density matrix ensemble. DQJ relies on sampling the jump times deterministically from an equally spaced jump time grid. The trajectories are then weighted by their probability to occur up to the final integration time $T$. 
In the following, we first provide the formal working principle of DQJ to first jump order, represented schematically in Fig.~\ref{fig:Method}(a) and (b). Then we explain how the required quantities emerge naturally from the numerical trajectory evolution. We summarize the DQJ algorithm to first order in pseudocode in Algorithm~\ref{alg:DQJ}.

The density matrix evolution at time $t\in [0,T]$ can be formally grouped by the number of jumps $n$ occurring up to the final integration time $T$ 

\begin{equation}
   \rho(t) = \sum_{n} p_n \rho^{(n)}(t)
\label{equ:JumpOrderDecomposition}
\end{equation}
where $p_n$ is the probability of a trajectory having $n$ jumps in the time interval $[0,T]$ and $\rho^{(n)}$ is the density matrix of all $n$-jump trajectories. For weakly dissipative systems, truncating this sum at low orders provides a good approximation of the density matrix.
The zeroth order density matrix $\rho^{(0)}(t)$ is simply the normalized evolution with Eq.~\eqref{equ:Heff} and no jump. The probability of the zero-jump trajectory is the probability of no jump occurring in the interval $[0,T]$, which we write as $p_0 = p^{(0)}_{[0,T]}$.

The first order contribution $p_1\rho^{(1)}(t)$ is given by an average over all jump trajectories $\rho^{(1)}_{\tau, \hat{L} }(t)$ weighted by the trajectory probability. A single jump trajectory consists of the normalized evolution with $H_\text{eff}$ to time $\tau\in [0,T]$, the application of a single quantum jump operator $\hat{L} \in J^1$, and a normalized propagation with $H_\text{eff}$ to the final time $ T$. Formally we write the first order as
\begin{equation}
\begin{aligned}
    p_1\rho^{(1)}(t)&= \sum_{\hat{L} \in J^1 }\int_{0}^{T}d\tau\; \tilde{p}_1(\tau, \hat{L} )\rho^{(1)}_{\tau, \hat{L} }(t)  \\
\end{aligned}
\label{equ:pn_rhon_integral}
\end{equation}
where $\tilde{p}_1(\tau, \hat{L} )$ is the trajectory probability density. 

DQJ approximates the integral in Eq.~\eqref{equ:pn_rhon_integral} with a sum over a deterministic, equally spaced grid of jump times $\Sigma^1 = \Delta t\cdot \{\frac{1}{2},\frac{3}{2} ,\dots, \left(N_\text{grid} - \frac{1}{2}\right)\}$ of spacing $\Delta t = T/N_\text{grid}$ with 
 
\begin{equation}
   p_1 \rho^{(1)}(t) \approx \sum_{\hat{L} \in J^1 ,\; \tau \in \Sigma^1} p_1(\tau,\hat{L}) \rho^{(1)}_{\tau, \hat{L}}(t) \quad .
\label{equ:pn_rhon_sum}
\end{equation}
This first order correction is depicted in Fig.~\ref{fig:Method}(a) and (b), schematically. Thus, we approximate the trajectory probability with a piecewise constant probability density, $p_1(\tau, \hat{L})=\Delta t\cdot \tilde{p}_1(\tau, \hat{L} )$, with the trajectory probability density given by

\begin{equation}
\begin{aligned}
\tilde{p}_1(\tau, \hat{L} ) =  p^{(0)}_{[0,\tau]} \; \langle \hat{L}^\dagger \hat{L} (\tau)\rangle
\end{aligned}
\label{equ:ProbabilityDensity}
\end{equation}
where $p^{(0)}_{[0,\tau]}$ is again the zero-jump probability in the interval $[0, \tau]$. In Eq.~\ref{equ:ProbabilityDensity}, we implicitly assume the probability of a second jump to  be zero in accordance to the first jump order expansion.

In a trace preserving manner, the density matrix up to first jump order can be defined by
\begin{equation}
   \rho(t) = p_0 \rho^{(0)}(t)+\frac{(1-p_0) }{\mathcal{N}_1}\left( p_1 \rho^{(1)}(t) \right)
\label{equ:rho_upto_1_sum}
\end{equation}
where the normalization is given by a sum over all single jump trajectory probabilities
\begin{equation}
\begin{aligned}
\mathcal{N}_1 &=\sum_{j \in J^1 ,\; \tau \in \Sigma^1} p_1(\tau, j) \quad .
\label{equ:Normalization1}
\end{aligned}
\end{equation}
This definition distributes all non-zero jump probability to the single jump trajectories, consistent with the approximation.

The representation of the density matrix in Eq.~\eqref{equ:rho_upto_1_sum} suggests the following algorithm, represented as pseudocode in Algorithm~\ref{alg:DQJ}. First, evolve the initial state with $H_\text{eff}$, generating the state $\ket{\psi(t)}_0$, which is in general not normalized. The probability of no jump having occurred until time $t$ is given by the squared norm of the unnormalized $\ket{\psi(t)}_0$ at that time, $p^{(0)}_{[0,t]} = \norm{\psi(t)_0}^2$. Hence, the probability of the zero jump trajectory is the squared norm at the final time, $p_0 =p^{(0)}_{[0,T]} $.
A single jump trajectory is depicted in Fig.~\ref{fig:Method}~(a). It is determined by first propagating the initial state to a jump time $\tau$ from the jump time grid $\Sigma^1$ with the effective Hamiltonian.  At time $\tau$ a jump operator $\hat{L}$ is applied to the state and the state is normalized again. Finally, the state is propagated to the final time $T$, again with the effective Hamiltonian, resulting in the state $\ket{\psi(T)}_{\tau, L}$. Further, the trajectory probability density Eq.~\eqref{equ:ProbabilityDensity} is simply given by the expectation value with respect to the unnormalized state before applying the jump operator, $\tilde{p}_1(\tau, \hat{L} ) = \bra{\psi(\tau)}\hat{L}^\dagger \hat{L}\ket{\psi(\tau)}_0$.

The trajectory $\rho^{(n)}_{\tau, \hat{L}}(t)$ for the zero and single jump trajectory is then the corresponding normed trajectory evolution
\begin{equation}
\begin{aligned}
\rho^{(1)}_{\tau, \hat{L}}(t) &=\frac{\ket{\psi(t)}\bra{\psi(t)}_{\tau, \hat{L}}}{\norm{\psi(t)_{\tau, \hat{L}}}^2}
\label{equ:rho_n_from_psi}
\end{aligned}
\end{equation}
which can be evaluated at the required time points, and for $t>\tau$. The sum of these first-order contributions generates $p_1\rho(1)(t)$, via Eq.~\eqref{equ:pn_rhon_sum}. For the first order DQJ method the number of such trajectories as a function of the grid discretization and the number of jump operators $N_J$ is given by $N_{\text{traj}, 1\text{ jump}} = 1+ N_{\text{grid}}  N_J$.

This concludes the DQJ method to first order. A pseudo-code summary is given in Algorithm~\ref{alg:DQJ}.

\begin{algorithm}[H]
    \caption{Single-jump DQJ Method}
    \label{alg:DQJ}
    \begin{algorithmic}[1]
        \Require Initial state $\ket{\psi(0)}$, effective Hamiltonian $H_{\text{eff}}$, jump operators $J = \{\hat{L}_i\}_{1\leq i\leq N_J}$, final time $T$,  midpoint jump time grid $\Sigma^1$
        \State $\ket{\psi(T)}\xleftarrow{H_{eff}}\,\ket{\psi(0)}$ \Comment{Zero-jump trajectory}
        \State \Call{RecordNorm}{$p_0 = \norm{\psi(T)}^2$}
        \For{$\tau \in \Sigma^1$} \Comment{Single jump time}
            \For{$\hat{L} \in J$} \Comment{Operator from jump operators}
                \State $\ket{\psi(\tau)}\xleftarrow{H_{eff}}\,\ket{\psi(0)}$  
                \State \Call{Record}{$\tilde{p}_1(\tau, \hat{L} ) = \bra{\psi(\tau)}\hat{L}^\dagger \hat{L}\ket{\psi(\tau)}$}
                \State $\ket{\psi(\tau)} \gets \hat{L}\ket{\psi(\tau)}\big/\sqrt{\bra{\psi(\tau)} \hat{L}^\dagger\hat{L} \ket{\psi(\tau)}}$ \Comment{Jump and normalize}
                \State $\ket{\psi(T)}\xleftarrow{H_{eff}}\,\ket{\psi(\tau)}$
            \EndFor
        \EndFor
        \State Normalize $\ket{\psi(t)}$ for all $t$ of interest
        \State Use Eq.~\eqref{equ:pn_rhon_sum}, \eqref{equ:rho_upto_1_sum}, and \eqref{equ:Normalization1} to calculate density matrix
    \end{algorithmic}
\end{algorithm}

\section{Two Jump DQJ}
\label{sec:Method_b}

The DQJ method to second order follows the same recipe as above. The jump order expansion  Eq.~\eqref{equ:JumpOrderDecomposition} for the integration interval $[0,T]$ is now truncated at second order, which requires the modification of the jump time grid and of the trajectory probabilities. Including the higher jump order allows for either capturing larger dissipative effects, increasing integration time, or increasing the accuracy to which the approximation is still applicable.

We first write out formally the contribution $p_2 \rho^{(2)}$ as an integral over the two jump times $\Vec{\tau} = \{\tau_1, \tau_2 \}$ at which the corresponding jump operators $\vec{L} = \{\hat{M}_1, \hat{M}_2\} \in J^2$ are applied
\begin{equation}
\begin{aligned}
    p_2\rho^{(2)}(t)&= \sum_{\vec{L} \in J^2 }\int_{0}^{T}d\tau_2\int_{0}^{\tau_2}d\tau_1\; \tilde{p}_2\left(\vec{\tau}, \vec{L} \right)\rho^{(2)}_{\vec{\tau}, \vec{L} }(t) \quad . \\
\end{aligned}
\label{equ:pn_rhon_integral2}
\end{equation}
Here, $J^2 $ denotes all ordered tuples of two jump operators from the list of jump operators $J$, e.g. $(L_3, L_2)\in J^2$. The integration domain respects that $\tau_1<\tau_2$, making the integration domain a simplex. To efficiently approximate the simplicial integral with a discrete sum
\begin{equation}
   p_2 \rho^{(2)}(t) \approx \sum_{\vec{L} \in J^2 ,\; \vec{\tau} \in \Sigma^2} p_2(\vec\tau,\vec{L}) \rho^{(2)}_{\vec{\tau}, \vec{L}}(t) \quad .
\label{equ:pn_rhon_sum_2}
\end{equation}
we sample over the Cartesian midpoints of the two dimensional jump time grid $\Sigma^1 \times \Sigma^1$ and additionally over the barycenter of the two-dimensional simplices for two closely spaced jumps 
\begin{equation}
\begin{aligned}
\Sigma^2 &= \Sigma^2_\text{cartesian} \cup \, \Sigma^2_\text{bary}  \\
\end{aligned}
\label{equ:2OrderJumpTimeGrid}
\end{equation}
with 
\begin{equation}
\begin{aligned}
\Sigma^2_\text{cartesian}&=\{(\tau_1,\tau_2)|\; \tau_1, \tau_2\in  \Sigma^1 ;  \;\tau_1<\tau_2\}\\[1em]
\Sigma^2_\text{bary}&= \left\{ (\tau - \Delta t/6,\; \tau+\Delta t/6)|\;\tau\in \Sigma^1   \right\}
\end{aligned}
\label{equ:JumpTimes2}
\end{equation}
as sketched in Fig.~\ref{fig:Method2D}. This choice of jump times allows for reusing parts of the single trajectory integrations.
The probability densities for the first and second jump order are defined as
\begin{equation}
\begin{aligned}
\tilde{p}_1\left(\tau, \hat{L} \right) &=  p^{(0)}_{[0,\tau]} \; \langle \hat{L}^\dagger \hat{L} (\tau)\rangle p^{(0)}_{[\tau,T]}\\
\tilde{p}_2\left(\vec{\tau}, \vec{L}\right) &=  p^{(0)}_{[0,\tau_1]} \; \langle \hat{M}_1^\dagger \hat{M}_1 (\tau_1)\rangle p^{(0)}_{[\tau_1,\tau_2]}\; \langle \hat{M}_2^\dagger \hat{M}_2 (\tau_2)\rangle
\end{aligned}
\label{equ:ProbabilityDensityM2N1}
\end{equation}
where the density for the single jump must now be corrected by the probability of no jump occurring up to the final integration time $p^{(0)}_{[\tau,T]}$, in contrast to Eq.~\eqref{equ:ProbabilityDensity}. Again, faithful to the approximation, we assume the probability of a third jump to be zero.

The trajectory probabilities are then approximated as piecewise constant probability densities, where the weight of the contribution depends on if the jumps $\vec{\tau}$ are elements of the cartesian or of the barycentric jump time grid
\begin{equation}
\begin{aligned}
\vec{\tau}\in\Sigma^2_\text{cartesian}&:\quad p_2\left(\vec{\tau}, \vec{L}\right) =  \left(\Delta t\right)^2 \cdot\tilde{p}_2\left(\vec{\tau}, \vec{L}\right)\\[1em]
\vec{\tau}\in\Sigma^2_\text{bary}&:\quad p_2\left(\vec{\tau}, \vec{L}\right) =  \frac{1}{2} \left(\Delta t\right)^2 \cdot\tilde{p}_2\left(\vec{\tau}, \vec{L}\right) \quad .
\end{aligned}
\label{equ:JumpTimes2Probabilities}
\end{equation}
The total trace preserving density matrix is defined as
\begin{equation}
   \rho(t) \approx p_0 \rho^{(0)}(t)+\frac{(1-p_0) }{\mathcal{N}_{1}+\mathcal{N}_{2}}\left(\sum_{n=1}^{2} p_n \rho^{(n)}(t) \right)
\label{equ:rho_upto_2_sum}
\end{equation}
with the normalization constants
\begin{equation}
\begin{aligned}
\mathcal{N}_n &=\sum_{\vec{L} \in J^n ,\; \vec{\tau} \in \Sigma^n} p_n(\vec{\tau}, \vec{L}) \quad .
\end{aligned}
\end{equation}

Numerically, all the above quantities for Eq.~\eqref{equ:rho_upto_2_sum}, using the approximation Eq~\eqref{equ:pn_rhon_sum} and \eqref{equ:pn_rhon_sum_2}, are accessible through the individual jump trajectories of the unnormalized $\ket{\psi(t)}_{\vec{\tau},\vec{L}}$.
The two-jump trajectories $\ket{\psi}_{\vec{\tau},\vec{L}}$ corresponding to  $\rho^{(2)}_{\vec{\tau}, \vec{L} }(t)$, with $\vec{L} =\{\hat{M}_1, \hat{M}_2\}$ and $\vec{\tau} =\{\tau_1, \tau_2\}$, are constructed similarly to the single trajectory. 
After the first jump $\hat{L}$ at time $\tau_1$ and subsequent normalization, the state is propagated to $\tau_2$ at which time one applies $\hat{M}_1$. The state is normalized and propagated to the final time. Hence, the probability density for a two-jump trajectory is $\tilde{p}_2\left(\vec{\tau}, \vec{L}\right)= \bra{\psi(\tau_1)}\hat{M}_1^\dagger \hat{M}_1\ket{\psi(\tau_1)}_0 \bra{\psi(\tau_2)}\hat{M}_2^\dagger \hat{M}_2\ket{\psi(\tau_2)}_{\tau_1, M_1}$, where the expectation values are with respect to the state before applying the jump operator. The single jump probability density is given by $\tilde{p}_1(\tau, \hat{L} ) = \bra{\psi(\tau)}\hat{L}^\dagger \hat{L}\ket{\psi(\tau)}_0\norm{\psi(T)_{\tau_1, L}}^2$, where the squared norm ensures for the probability that no jump appears after the first jump.
The individual trajectories are then normalized at the times $t$ for which the density matrix should be evaluated as in Eq~\eqref{equ:rho_n_from_psi}. For the second order DQJ method, the total number of such trajectories as a function of the grid discretization and the number of jump operators $N_J$ is given by $N_{\text{traj}, 2 \text{ jumps}} = 1+ N_{\text{grid}} N_J + \frac{1}{2}N_{\text{grid}}\left(N_{\text{grid}}+1\right)N_J^2$.

\section{Applicable Regime and  Error scaling}
\label{sec:Method_c}

The DQJ method outperforms SQJ when dissipation is weak and trajectories with few jumps dominate the ensemble of trajectories. It then suffices to only calculate these few jump orders. DQJ makes use of the fact that it is more efficient to deterministically sample the jump times from an equally spaced jump time grid than to sample randomly from the jump time probability distribution. The error of the DQJ method can then be thought of as the Riemann sum midpoint error of the integral jump time integral Eq.~\eqref{equ:pn_rhon_integral} and \eqref{equ:pn_rhon_integral2}.
For finite dissipation, the scaling of the error in the number of trajectories $N_\text{traj}$ can then be divided into two domains (Fig.~\ref{fig:Method}~(c)). Once the jump time grid $\Sigma^1$ is fine enough to capture the maximal frequency $f_{\max}$ of the density matrix evolution, $1/(\Delta t)<f_{\max}$, we enter the asymptotic domain. In this domain the individual jump orders $p_n \rho^{(n)}$ are determined ever more exactly and follow an asymptotic scaling behavior. For maximal jump order $n$ considered in the DQJ method,  the error on a single matrix element of the density matrix $\rho_{ij}$ scales with 
\begin{equation}
\begin{aligned}
\mathcal{E}^{(\mathrm{DQJ})}[\rho_{ij}(T)] \propto \frac{1}{\left(N_\text{traj}\right)^{2/n}} \quad.
\end{aligned}
\label{equ:ProptoScaling}
\end{equation}
For up to order $n=3$, this outperforms the SQJ method.
A detailed error bound is derived in Appendix~\ref{app:Error}. 

This scaling holds for all $\rho_{ij}(t)$ at all discrete times $t\in \Delta t\cdot\{1,2,\dots, N_\text{grid}\}$, which includes the final time $T$. At intermediate times, due to a discontinuity in the jump time integral, the scaling reduces to $1/\left(N_\text{traj}\right)^{1/n}$. Further, we note that for frequency space observables which are a function of $\rho(\omega)$, the error in the DQJ methods scales favorably compared to the SQJ method, as is shown in the example Sec.~\ref{subsec:Kerr-oscillator}.

As metric for the quantitative performance we introduce the fidelity
\begin{equation}
    F\left(\sigma, \rho\right) = \left(\Tr\left[\sqrt{\sqrt{\sigma}\rho\sqrt{\sigma}}\right]\right)^2\quad .
\label{equ:fidelity}
\end{equation}
With this, the target infidelity scales with (Appendix~\ref{app:Error})
\begin{equation}
    1-F\left(\sigma, \rho\right) \propto \frac{1}{\left(N_\text{traj}\right)^{4/n}} \quad .
\label{equ:InfidScaling}
\end{equation}
We note that this and related quantities are widely used in the context of quantum technology, providing a key motivation for our method in general, and this metric in particular. At an accuracy level at which the higher neglected jump orders become relevant, the error scaling in the number of trajectories plateaus. Although each jump order contribution continues to scale with the above asymptotic scaling, further trajectories do not enhance the approximation of the total density matrix. An upper bound of the plateau height can be estimated by determining the maximal probability of more jumps occurring in the time evolution than the jump order considered in the DQJ method. Assuming a simplified Poissonian process, to leading order in the effective jump rate the error plateau is given by $\Pr(\# \text{jumps}>n) = \left(\gamma_{\text{eff}}\,T\right)^{n+1}/(n+1)!$. For fidelity estimates of full rank density matrices, the probability must be squared to find the the plateau height.

\section{Examples}
\label{sec:Examples}

We demonstrate the DQJ method for two quantum systems with weak dissipation and compare its performance to the SQJ method. We compare both methods to the evolution of the full density matrix propagated with the Lindblad Master equation. The numerical simulations recover the asymptotic scaling domain discussed in Sec.~\ref{sec:Method_c} and clearly illustrate the regime in which our DQJ method outperforms the SQJ method.

For a meaningful comparison, we employ an adapted version of the SQJ method (Appendix~\ref{app:SQJ}), which respects the no-jump probability by constraining the range of the randomly sampled number. In contrast, applying SQJ naively would require on the order of $ 1/(\gamma T)$ more trajectories, where $\gamma$ is a characteristic decay rate of the system.

\subsection{Transverse-field Ising Model}
We consider the transverse-field Ising model (TFIM) to demonstrate the efficiency of the DQJ method to simulate large dissipative systems.
The 1D TFIM consists of a chain of $N$ nearest neighbor coupled two-level systems. 
\begin{equation}
    H_{\text{TFIM}} = g \sum_k \sigma^x_k + J \sum_{\langle k\,l\rangle } \sigma^z_k\sigma^z_l
\end{equation}
We introduce weak dissipation in terms of local amplitude damping $\sigma^-$
\begin{equation}
    J^1_{\text{TFIM}} = \sqrt{\gamma} \cdot\{\sigma^-_1,\sigma^-_2, \dots, \sigma^-_N\}
\end{equation}
where each entry corresponds to an element $L_j = \sqrt{\gamma} \sigma^-_j$ in Sec.~\ref{sec:Method_a}. The state $\ket{\psi} = \ket{\uparrow\uparrow\dots\uparrow}$ is propagated to the final time $T$ using DQJ, SQJ and the Lindblad master equation
\begin{equation}
    \mathcal{L}[\rho] = -i[H, \rho] + \sum_{\hat{L}\in J^1} \hat{L}\rho \hat{L}^\dagger - \frac{1}{2}\{\hat{L}^\dagger \hat{L}, \rho\}
\end{equation}
with $\rho_0 = \ket{\psi}\bra{\psi} $. We work with unitless quantities, such that $T=1$, $J=\pi/2$, and $g/J = 2 \pi$. 
\begin{figure}[ht]
    \centering
    \includegraphics[width=\columnwidth]{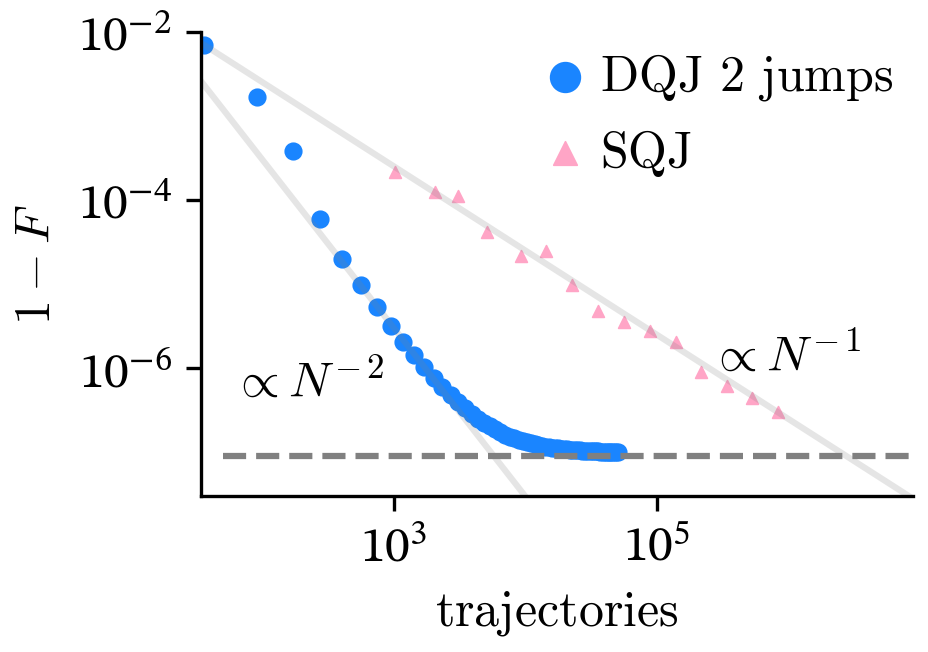} 
    \caption{Infidelity of the density matrix calculated with the DQJ (blue) and the modified SQJ (red) method with the true density matrix evolution. The infidelity is evaluated at time $T=1$ for a $5$ qubit TFIM system with $\gamma\, T  \approx 0.03$. Numerically, we recover the asymptotic scaling of Eq.~\eqref{equ:InfidScaling}. The scaling functions (grey) serve as a guide to the eye. The minimal infidelity is set by the contribution of trajectories with more than two jumps.  }
    \label{fig:ExampleFTIM}
\end{figure}

In Fig.~\ref{fig:ExampleFTIM} we compare the fidelity, given in Eq.~\eqref{equ:fidelity}, of the density matrix at time $T$ calculated via the two approximate methods with the density matrix of the full density matrix using the master equation, $F(\rho_\mathcal{L}, \rho_{\text{DQJ}})$ and $F(\rho_\mathcal{L}, \rho_{\text{SQJ}})$. We have set $\gamma = 0.03$. The asymptotic regime of the DQJ method at the 2-jump level scales with with $1/N^2$, as discussed in Sec.~\ref{sec:Method_c}. 

This regime plateaus at a fidelity of about $1.0\cdot 10^{-7}$. This is close to the Poissonian estimate of $\left(\Pr(\# \text{jumps}>2)\right)^2=3.1\cdot10^{-7}$, as described in Sec.~\ref{sec:Method_c}, with an effective maximal decay rate of $\gamma_{\text{eff}} \leq \gamma \cdot N_{\text{qubits}}$. 

In comparison, the error of the modified SQJ method scales more slowly and homogeneously with $1/N$ over all trajectory numbers . To reach the same target infidelity, the SQJ method requires upto several orders of magnitude more trajectories than our DQJ method. 

For weaker dissipation the plateau shifts to lower infidelities and the asymptotic scaling domain is prolonged. In Appendix~\ref{app:TFIM}, we further show the scaling of DQJ in for different dissipation strength $\gamma$ and qubit numbers for the single and two jump DQJ method. We note for completeness, that the asymptotic scaling must not always be visible in scaling plots similar to Fig.~\ref{fig:ExampleFTIM}. Namely, if the density matrix has high frequency components, the onset of the asymptotic domain might already be in the plateaued domain. The scaling then appears even more drastic as the error collapses to the error floor once the grid captures the highest  frequency component of the evolution.

\subsection{Kerr-oscillator}
\label{subsec:Kerr-oscillator}

\begin{figure}[ht]
    \centering
    \includegraphics[width=\columnwidth]{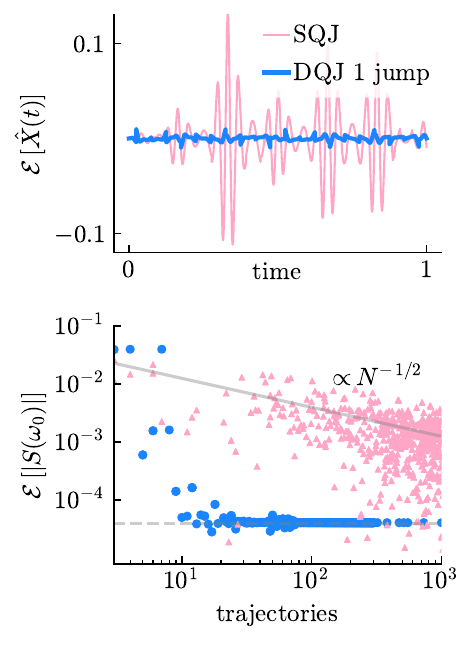}  
    \caption{Top: Error of the $\langle\hat{X}\rangle$ evolution in the  Kerr-oscillator. The DQJ and the SQJ method are compared to the full density matrix Lindbladian simulation. A total of $21$ trajectories were propagated for both methods. Bottom: Error scaling in the number of trajectories for the $X$-quadrature spectrum at frequency $\omega_0$. for the SQJ (red) and DQJ (blue) method compared to the true full density evolution. The dissipation strength is set to $\gamma\, T  = 0.32$.  The scaling functions (grey) in the number of trajectories for the SQJ is shown to guide the eye.}
    \label{fig:ExampleKerr}
\end{figure}

Anharmonic oscillators are ubiquitous systems in quantum mechanics. They are found in cavity and circuit QED setups and plays an integral role in many quantum computing protocols. It is therefore of importance to determine its time evolution with high precision, when evaluating quantum computation operations. The Kerr-oscillator Hamiltonian is
\begin{equation}
    H = \omega_0 \, \hat{n} + \frac{1}{2}  \omega_K \, \hat{n}\left(\hat{n}-1\right)
\end{equation}
where $\omega_0$ is the bare frequency of the harmonic oscillator and $\omega_K$ determines the magnitude of the anharmonicity.  As an exemplary observable we measure the quadrature
\begin{equation}
    \hat{X} = \frac{1}{\sqrt{2}} \left( a + a^\dagger \right) 
\end{equation}
and calculate the spectrum of its evolution as
\begin{equation}
    S(\omega) = \int_0^T dt \; \langle\hat{X}(t)\rangle e^{-i\omega t}
\end{equation}
We calculate $\abs{S(\omega_0)}$ with both the SQJ and our DQJ method and determine the error $\mathcal{E}[\abs{S(\omega_0)}] = \abs{S_\text{method}(\omega_0)}-\abs{S_\mathcal{L}(\omega_0)}$ to the true value computed with the full density matrix evolution. For the simulation we choose $T = 1$, $\omega = 24 \pi $, $\omega_K =\omega/4$ and start in a coherent state $\ket{\alpha} \approx 1.1$, such that $\langle\hat{X}\rangle = 1.5$. In Fig.~\ref{fig:ExampleKerr}(a), we first show the error $\mathcal{E}[\hat{X}(t)] = \langle \hat{X}(t)\rangle-\langle \hat{X}(t)\rangle_{\mathcal L}$ of the time evolution of $\langle \hat{X}(t)\rangle$, for the SQJ method and the DQJ method, and the 1-jump level. In Fig.~\ref{fig:ExampleKerr}(b), we show the error $\mathcal{E}[\abs{S(\omega_0)}] = \abs{S(\omega_0)}-\abs{S(\omega_0)}_{\mathcal L}$  for the two methods, as a function of the trajectories. We take into account Fock states up to $n_\text{Fock} = 6$.

This error plateau appears at $\mathcal{E}[\abs{S(\omega_0)}] \approx 4.1\cdot 10^{-5}$. The small infidelity can be reached even with the first jump DQJ method because the jump probability is reduced after the first jump. In contrast, the SQJ method shows a much higher variability of the resulting error due to the stochastic sampling method. The scaling function of the SQJ method crosses the error plateau at $\approx 10^6$ trajectories, highlighting the efficiency with which time integrated quantities are determined using DQJ.

\section{Conclusion}
\label{sec:Conclusion}

We have proposed a quantum jump algorithm for weakly dissipative systems, the deterministic quantum jump (DQJ) method. This method utilizes a few-jump expansion at deterministic times, given by a time grid. It thus differs from the standard quantum jump (SQJ) methods, which utilizes jumps at random times. The deterministic jump method outperforms these methods in the weakly dissipative limit, because it eliminates the stochastic error generated by the random times. The target infidelity of the reconstructed density matrix of our proposed method displays a power-law scaling with the number of trajectories $N$, that outperforms the SQJ methods, in particular at the single-jump and two-jump level. It is limited, however, by the truncation error due to the truncation at these levels. This limitation is negligible compared to other error sources, exactly in the weakly-dissipative regime, in which quantum jumps are rare, which motives the main regime of operation of our method. We demonstrate our method for two examples, dynamics of the transverse-field Ising model and of a Kerr oscillator, exemplifying its performance. We have proposed this method motivated by the simulation and development of quantum technology, which is intrinsically focused on weakly-dissipative systems, in which quantum jumps are necessarily rare events.

\section*{Acknowledgements} 
\label{sec:Acknowledgemnts}
The project is financed by ERDF of the European Union and by ’Fonds of the Hamburg Ministry of Science, Research, Equalities and Districts (BWFGB)’. We acknowledge funding by the Cluster of Excellence “Advanced Imaging of Matter” (EXC 2056) Project No.390715994.

\newpage

\bibliography{DQJ_Method}

\appendix
\section{Error scaling} 
\label{app:Error}

Here we derive the dominant error scaling of the DQJ method and and provide an upper bound to the error of the method. 

The exact formal first jump order contribution is given by
\begin{equation}
\begin{aligned}
    p_1\rho^{(1)}(t)&= \int_{0}^{T}d\tau\; \tilde{p}_1(\tau )\rho^{(1)}_{\tau }(t)  \\
\end{aligned}
\label{equ_app:pn_rhon_integral}
\end{equation}
where we have assumed only a single jump operator $\hat{L}$ acts on the system. For $t=T$, Eq.~\eqref{equ_app:pn_rhon_integral} is an integral over smooth and continuous function.

The DQJ method  approximates this integral over the jump times $\tau$ by sampling them from the equally spaced jump time grid. Here, $\Sigma^1$ contains the midpoints of the intervals of size $\Delta t$. It therefore serves as an integration similar to the Riemann sum of a one dimensional integral. It hence acquires the same error scaling in the grid discretization

\begin{equation}
\begin{aligned}
    \abs{\int_{0}^{T}d\tau\, \tilde{p}_1(\tau )\rho^{(1)}_{\tau }(t)  - \sum_{ \tau \in \Sigma^1} p_1(\tau) \rho^{(1)}_{\tau}(t) }\leq \frac{\norm{ \partial_{\tau \tau}\left( \tilde{p}_\tau\rho_{\tau }\right)} T^3}{24 N_\text{grid}^2}
\end{aligned}
\label{equ_app:ErrorRiemann}
\end{equation}
where the second derivative is with respect to the jump time $\tau $ at propagation time $t$. The norm is to be read as supremum over all $\tau$.
From Eq.~\eqref{equ_app:ErrorRiemann} and the number of total trajectories for the DQJ method for each order up to order three 
\begin{equation}
\begin{aligned}
N_{ 1\text{jump}} &= 1+N_{\text{grid}}N_J \\
N_{2 \text{jumps}} &=  N_{ 1\text{jump}} + \frac{1}{2}N_{\text{grid}}\left(N_{\text{grid}}+1\right)N_J^2 \\
N_{3 \text{jumps}} &= N_{2 \text{jumps}} + N_{\text{grid}}\left(N_{\text{grid}}^2-2N_{\text{grid}} + 2\right)N_J^3\\
\end{aligned}
\label{equ:NgridToNTraj}
\end{equation}
we can immediately see that the dominant scaling in the number of trajectories for a matrix entry must follow the $\propto 1/N^{2/n}$ discussed in Eq.~\eqref{equ:ProptoScaling}.

We now determine the prefactor of this scaling and therefore an explicit upper bound on the error.

Using the superoperator
\begin{equation}
    \mathcal{L}[\rho] = -i \left(H_\text{eff}\rho-\rho H_\text{eff}^\dagger\right)
\end{equation}
the time evolution is given by
\begin{equation}
    \mathcal{U}_\tau = e^{\tau \mathcal L}
\end{equation}
A general property of the superoperator is
\begin{equation}
\begin{aligned}
        \partial_\tau\mathcal{U}_\tau = \mathcal{L}\;e^{\tau \mathcal L}
\end{aligned}
\end{equation}
To describe a jump trajectory, we have to include the jump as well, we therefore define a jump superoperator $\mathcal{J}$ with

\begin{equation}
    \mathcal{J}[\rho]  = \frac{L \rho L^\dagger}{\Tr{L \rho L^\dagger}}
\end{equation}

\subsubsection*{Density matrix derivatives}
In the following we omit the $t$ dependence for clarity and write $\rho_\tau \equiv \rho_\tau^{(1)}$.
We can then write the normalized density matrix  $\rho_\tau$ as
\begin{equation}
\rho_\tau
=
  \begin{cases}
    \dfrac{1}{p^{(0)}_{[0,t]}}\,\mathcal{U}_{t}[\rho_0], 
      & \tau \geq t,\\[1ex]
    \dfrac{1}{p^{(0)}_{[\tau, t]}}\,\mathcal{U}_{t-\tau}\,\mathcal{J}\,\mathcal{U}_{\tau}[\rho_0], 
      & \tau < t.
  \end{cases}
\end{equation}
Taking the first and second derivative
 \begin{equation}
 \begin{aligned}
    \partial_\tau\rho_{\tau> t }&=  0\\
    \partial_{\tau \tau}\rho_{\tau> t }&= 0\\
\end{aligned}
\end{equation}
Using $\rho = \dfrac{1}{z}A$ with $A = \mathcal{U}_{t-\tau}\,\mathcal{J}\,\mathcal{U}_{\tau}$ and $z = p^{(0)}_{[\tau, t]}$
\begin{equation}
\begin{aligned}
\partial_\tau\rho_{\tau\leq t }&=  \left(\frac{1}{z}\left(\partial_\tau A \right)
    - \dfrac{\left(\partial_\tau z\right)}{z^2}A\right)[\rho_0]\\
\partial_{\tau \tau}\rho_{\tau\leq t }&=\bigg(\bigg(2\dfrac{\left(\partial_{ \tau} z\right)^2}{z^3}-\dfrac{\left(\partial_{\tau \tau} z\right)}{z^2}\bigg)A\\
&\quad-2\dfrac{\left(\partial_\tau z\right)}{z^2}\left(\partial_\tau A \right) + \frac{1}{z}\left(\partial_{\tau\tau} A  
\right)\bigg)[\rho_0]
\end{aligned}
\end{equation}
Using
\begin{equation}
\begin{aligned}
    \partial_{\tau}A &=  -\mathcal{L}\,\mathcal{U}_{t-\tau}\mathcal{J}\,\mathcal{U}_{\tau} + \mathcal{U}_{t-\tau}\mathcal{J}\,\mathcal{L}\,\mathcal{U}_{\tau} \\
    &= \mathcal{U}_{t-\tau} [\mathcal{J}, \mathcal{L}]\,\mathcal{U}_{\tau}\\
\end{aligned}
\end{equation}
\begin{equation}
\begin{aligned}
        \partial_{\tau \tau}A &=  \mathcal{L}^2\,\mathcal{U}_{t-\tau}\mathcal{J}\,\mathcal{U}_{\tau} - 2\mathcal{L}\,\mathcal{U}_{t-\tau}\mathcal{J}\,\mathcal{L}\,\mathcal{U}_{\tau}+ \mathcal{U}_{t-\tau}\mathcal{J}\,\mathcal{L}^2\,\mathcal{U}_{\tau} \\
    &=\mathcal{U}_{t-\tau} [[\mathcal{J}, \mathcal{L}],\mathcal{L}]\,\mathcal{U}_{\tau}
\end{aligned}
\end{equation}
and
\begin{equation}
\begin{aligned}
     \partial_{\tau }z &=\langle L^\dagger L(\tau)\rangle\; z\\
     \partial_{\tau \tau }z &= \left(\langle L^\dagger L(\tau)\rangle^2 +\partial_\tau\langle L^\dagger L(\tau)\rangle \right)\; z
\end{aligned}
\end{equation}

\subsubsection*{Probability density derivatives}
The probability density $\tilde{p}_\tau $ of the first jump density matrix is given, using $y \equiv  p^{(0)}_{[0,\tau]} \leq 1 $, by
\begin{equation}
\begin{aligned}
\tilde{p}_\tau & = \langle L^\dagger L (\tau)\rangle \; y = -\partial_\tau y\\
\partial_\tau\tilde{p}_\tau & = \left(\partial_\tau\langle L^\dagger L (\tau) \rangle  - \langle L^\dagger L (\tau) \rangle^2 \right)y\\
\partial_{\tau\tau}\tilde{p}_{\tau}&=\big(\partial_{\tau\tau}\langle L^\dagger L (\tau) \rangle  \\
&\quad-3 \langle L^\dagger L (\tau) \rangle \left(\partial_\tau \langle L^\dagger L (\tau) \rangle\right) + \langle L^\dagger L (\tau) \rangle^3 \big)\,y
\end{aligned}
\end{equation}

\subsubsection*{Bounds}
Using $\norm{\mathcal{U}} \leq 1$, $\norm{\mathcal{J}} = 1 $, $\abs{y}\leq 1$, $\norm{A}\leq 1$, and
\begin{equation}
\begin{aligned}
\norm{\mathcal{L}} &\leq 2\norm{H_\text{eff}} +\norm{\sum_k L^\dagger_k L_k} \\\
\norm{\partial_\tau A} &\leq 2\norm{\mathcal{L}}\\
\norm{\partial_{\tau \tau} A} &\leq 4\norm{\mathcal{L}}^2\\
\end{aligned}
\end{equation}
Defining
\begin{equation}
\begin{aligned}
\abs{\langle L^\dagger L (\tau)\rangle}  &\leq l_0\\
\abs{\partial_\tau\langle L^\dagger L (\tau)\rangle}  &\leq l_1 \leq l_0 \norm{\mathcal{L}}  \\
\abs{\partial_{\tau \tau}\langle L^\dagger L (\tau)\rangle}  &\leq l_2 \leq l_0 \norm{\mathcal{L}} ^2\\
\end{aligned}
\end{equation}
We find
\begin{equation}
\begin{aligned}
    z_{\text{min}} \leq \abs{z}&\leq 1 \\
     \abs{\partial_{\tau }z} &\leq l_0\\
     \abs{\partial_{\tau \tau }z} &\leq l_0^2 + l_1 \leq l_0^2 + l_0 \norm{\mathcal{L}} 
\end{aligned}
\end{equation}
Then the density matrix derivatives are bounded by
\begin{equation}
\begin{aligned}
\norm{\rho_\tau}&= 1\\
    \norm{\partial_{\tau }\rho_\tau}&\leq \dfrac{1}{z_\text{min}}\left(2\norm{\mathcal{L}} + \frac{l_0}{z_\text{min}}\right)\\
    \norm{\partial_{\tau \tau}\rho_\tau}&\leq \dfrac{1}{z_\text{min}}\left(4\norm{\mathcal{L}}^2 + \frac{4 l_0}{z_\text{min}}\norm{\mathcal{L}} +\frac{l_0^2 + l_1}{z_\text{min}} +\frac{2 l_0^2}{z_\text{min}^2}  \right)\\
\end{aligned}
\end{equation}
or in terms of the $\norm{\mathcal{L}}$
\begin{equation}
\begin{aligned}
    \norm{\partial_{\tau \tau}\rho_\tau}&\leq \dfrac{1}{z_\text{min}}\left(4\norm{\mathcal{L}}^2 + \frac{4 l_0}{z_\text{min}}\norm{\mathcal{L}} +\frac{l_0^2 + l_0 \norm{\mathcal{L}} }{z_\text{min}} +\frac{2 l_0^2}{z_\text{min}^2}  \right)\\
\end{aligned}
\end{equation}
The probability density is bounded by
\begin{equation}
\begin{aligned}
\norm{\tilde{p}_\tau }& \leq l_0\\
\norm{\partial_\tau\tilde{p}_\tau} & \leq l_1 + l_0^2\\
\norm{\partial_{\tau\tau}\tilde{p}_{\tau}}&\leq l_2 + 3l_0 l_1 + l_0^3
\end{aligned}
\end{equation}
or in terms of the $\norm{\mathcal{L}}$
\begin{equation}
\begin{aligned}
\norm{\partial_\tau\tilde{p}_\tau} & \leq l_0^2 + l_0 \norm{\mathcal{L}}\\
\norm{\partial_{\tau\tau}\tilde{p}_{\tau}}&\leq l_0 \norm{\mathcal{L}}^2 + 3l_0^2\norm{\mathcal{L}} + l_0^3
\end{aligned}
\end{equation}

\begin{figure}[!htbp]
    \centering
    \includegraphics[width=\linewidth]{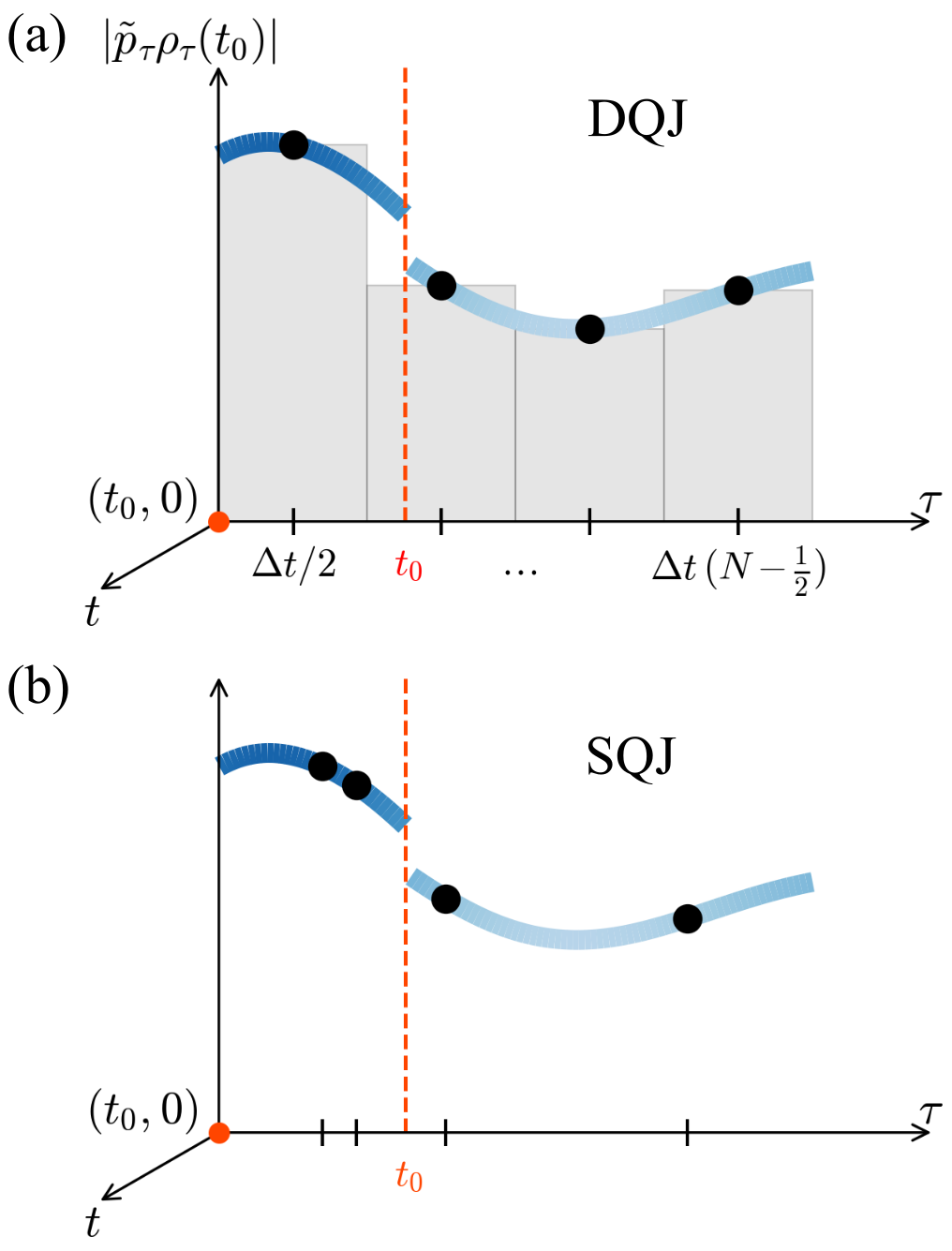} 
    \caption{(a) Schematic jump time integral of Eq.~\eqref{equ:pn_rhon_integral} at time $t = t_0$ and the sampled jump time trajectories (black dots) used in the DQJ approximation, Eq.~\eqref{equ:pn_rhon_sum}. 
    (b) In SQJ, the integral is instead sampled with random points from the $\tilde{p}_\tau$ trajectory probability distribution. The jump time integral has a discontinuity (red) at the jump time $\tau = t_0$ for the time $t= t_0$.}
    \label{fig:Method1D}
\end{figure}

\begin{figure}[!htbp]
    \centering
    \includegraphics[width=\linewidth]{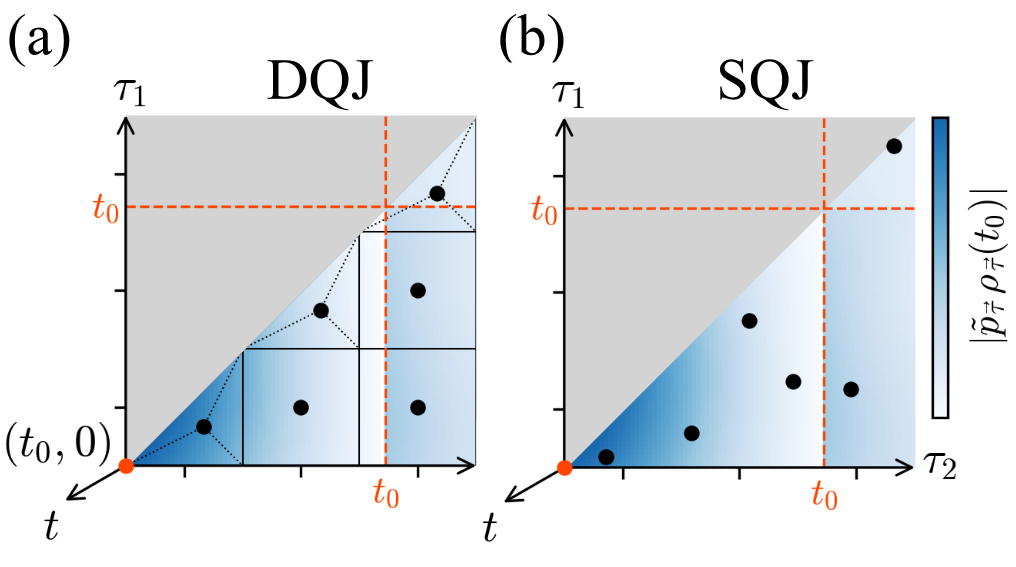}  
    \caption{(a) Two-time jump time integral Eq.~\eqref{equ:pn_rhon_integral2}. DQJ integrates over the simplex spanned by $\tau_1 <\tau_2$ by sampling the jump times deterministically over the cartesian midpoint grid $\Sigma^2_\text{cartesian}$ and the barycenters $ \Sigma^2_\text{bary}$ for $\tau_1 \lesssim \tau_2$. 
    (b) SQJ again samples randomly from the trajectory probability distribution $\tilde{p}_2\left(\vec{\tau}\right)$. }
    \label{fig:Method2D}
\end{figure}

\subsubsection*{Final expression for error bound}
The full argument becomes
\begin{equation}
\begin{aligned}
     \partial_{\tau \tau}\left( \tilde{p}_\tau\rho_{\tau }\right) &= \left(\partial_{\tau \tau}\tilde{p}_\tau\right)\rho_\tau + \tilde{p}_\tau\left(\partial_{\tau \tau}\rho_\tau\right) \\
    &\quad+ 2\left(\partial_{\tau }\tilde{p}_\tau\right)\left(\partial_{\tau }\rho_\tau\right)
\end{aligned}
\end{equation}
then assuming a low jump rate, we only keep the terms proportional to $l_0 \propto \gamma$ and find
\begin{equation}
\norm{ \partial_{\tau \tau}\left( \tilde{p}_\tau\rho_{\tau }\right)} \lessapprox 
  \begin{cases}
    l_0 \norm{\mathcal{L}}^2,  
      & \tau \geq t,\\[1ex]
    l_0 \norm{\mathcal{L}}^2 \left(1+\frac{8}{z_\text{min}}\right)& \tau < t
  \end{cases}
\end{equation}
Hence the error for the estimate of the density matrix up to first jump order and assuming for small loss rates $1/z_\text{min} \approx 1$ is given by
\begin{equation}
    \begin{aligned}
        \norm{\rho(T) - (p_0\rho^{(0)}(T) + p_1\rho^{(1)}(T))}\leq (1-p_0)\frac{3l_0 \norm{\mathcal{L}}^2T^3}{8 N_{\text{grid}}^2}
    \end{aligned}
\label{equ:RiemannError}
\end{equation}

\subsubsection{Fidelity error scaling}
\label{app_sub:Fidelity}

The Infidelity can be written in terms of the Bures distance as 
\begin{equation}
\begin{aligned}
        1-F(\rho, \sigma) =  d_B^2 \quad .
\end{aligned}
\end{equation}
A small perturbation to the density matrix leads to the Bures distance to change as \cite{hubnerExplicitComputationBures1992}
\begin{equation}
\begin{aligned}
        d_B(\rho, d\rho)^2 = \sum_{ij}\frac{1}{2}\frac{\abs{\bra{i}d\rho\ket{j}}^2}{\lambda_i + \lambda_j}
\end{aligned}
\end{equation}
where it is assumed that the density matrix is of full rank so that no eigenvalue is zero. This assumption is indeed the most generic case for simulating coupling to the environment in the weakly dissipative regime. If the density matrix is not of full rank it can either be because jump operators do not act on a subspace of the Hilbert space. It would then suffice to propagate the state with the Schroedinger equation only. The density matrix could also lose rank in the long time limit if the ground state is a pure state attractor. Because of the DQJ regime with $\gamma T\ll1$ this purification is not relevant to our consideration.

We then find that if $d\rho \propto 1/N^2$, see Eq.~\eqref{equ:RiemannError}, to lowest order in $d\rho$
\begin{equation}
\begin{aligned}
        1-F(\rho, \sigma) \propto \frac{1}{N^4}
\end{aligned}
\end{equation}

\section{TFIM}
\label{app:TFIM}

\begin{figure}[!htbp]
    \centering
    \includegraphics[width=\linewidth]{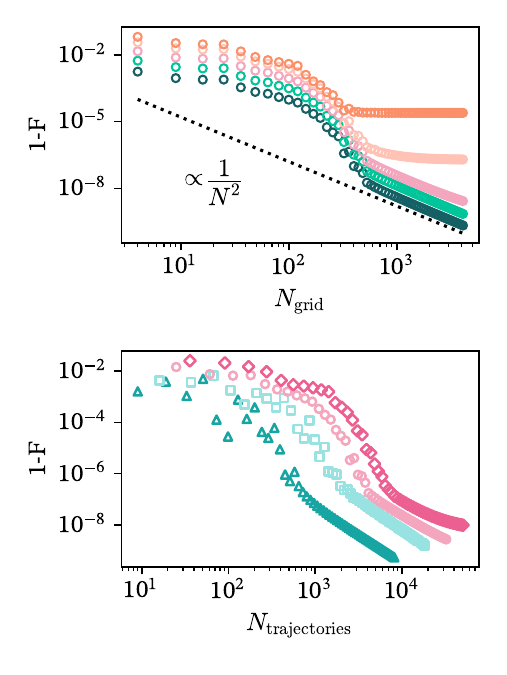}
    \caption{Top: 2 jump DQJ scaling of the $4$ qubit Transverse Field Ising Model of infidelity with the grid discretization. From dark green to orange the dissipation strength is varied with $\gamma = 0.003$, $0.01$, $0.03$, $0.1$,  $0.3$, where $\omega = 4\pi$ for $T=1$ is kept constant. At around $\Delta t \approx 1/300$ the asymptotic scaling sets in. The dotted black line serves as a visual guide.
    Bottom: Increasing the number of qubits from $2-5$ (green to red) for $\gamma = 0.03$. For more qubits the effective $\gamma_\text{eff}$ increases. Further, due to the scaling of the number of trajectories with $N_J^2$ the scaling curve shifts to more trajectories. }
    \label{fig:TFIMScaling}
\end{figure}

\section{Third order DQJ} 
\label{app:DQJ3}
In this section, we provide an efficient jump time grid $\Sigma^3$, which maximizes the number of jump times on the cartesian grid. This allows to share parts of the time evolution between trajectories, possibly reducing the number of trajectories by a factor.
The third order jump time grid is given by
\begin{equation}
\begin{aligned}
\Sigma^3 &= \Sigma^3_\text{cartesian} \cup \; \Sigma^3_\text{bary}   \\
\end{aligned}
\label{equ:3OrderJumpTimeGrid}
\end{equation}
with 
\begin{equation}
\begin{aligned}
\Sigma^3_\text{cartesian}&=\{(\tau_1,\tau_2, \tau_3)|\; \tau_1, \tau_2, \tau_3\in  \Sigma^1 ;  \;\tau_1<\tau_2<\tau_3\}\\[1em]
\Sigma^3_\text{bary}&=\Sigma^3_\text{A} \cup \Sigma^3_\text{B}\cup \Sigma^3_\text{C} \;.
\end{aligned}
\end{equation}
Here $\Sigma^3_\text{A/B/C} $ refer to the three different types of edges that can appear which are not entirely covered by the cartesian. 
\begin{equation}
\begin{aligned}
\Sigma^3_\text{A}&=\{\left(\tau-\frac{\Delta t}{4},\tau, \tau+\frac{\Delta t}{4}\right)|\; \tau \in  \Sigma^1 \}\\[1em]
\Sigma^3_\text{B}&=\{\Delta t\left(\frac{1}{2}, \frac{1}{3},\frac{2}{3}\right) + k \,\Delta t\left(0,1,1\right) + l\,\Delta t\left(1, 0, 0\right)|\\
&\quad\; 0\leq l < k<N_{\text{grid}} \}\\
\Sigma^3_\text{C}&=\{\Delta t\left(\frac{1}{3}, \frac{2}{3},\frac{1}{2}\right) + k\,\Delta t \left(0,0,1\right) + l\,\Delta t\left(1, 1, 0\right)|\\
&\quad\; 0\leq l < k<N_{\text{grid}} \}\\
\end{aligned}
\end{equation}

The trajectory probabilities are then given by 

\begin{equation}
\begin{aligned}
\tilde{p}_1\left(\tau, \hat{L} \right) &=  p^{(0)}_{[0,\tau]} \; \langle \hat{L}^\dagger \hat{L} (\tau)\rangle p^{(0)}_{[\tau,T]}\\
\tilde{p}_2\left(\vec{\tau}, \vec{L}\right) &=  p^{(0)}_{[0,\tau_1]} \; \langle \hat{M}_1^\dagger \hat{M}_1 (\tau_1)\rangle p^{(0)}_{[\tau_1,\tau_2]}\; \langle \hat{M}_2^\dagger \hat{M}_2 (\tau_2)\rangle p^{(0)}_{[\tau_2,T]}\\\
\tilde{p}_3\left(\vec{\tau}, \vec{L}\right) &=  p^{(0)}_{[0,\tau_1]} \; \langle \hat{M}_1^\dagger \hat{M}_1 (\tau_1)\rangle p^{(0)}_{[\tau_1,\tau_2]}\; \langle \hat{M}_2^\dagger \hat{M}_2 (\tau_2)\rangle p^{(0)}_{[\tau_2,T]}\\&\quad\langle \hat{M}_3^\dagger \hat{M}_3 (\tau_3)\rangle 
\end{aligned}
\label{equ:ProbabilityDensityM3N1}
\end{equation}

\begin{algorithm}[H]
    \caption{Adapted SQJ Method}
    \label{alg:SQJ}
    \begin{algorithmic}[1]
        \Require Initial state $\ket{\psi(0)}$, effective Hamiltonian $H_{\text{eff}}$, jump operators $J^1 = \{\hat{L}_i\}_{1\leq i\leq N_J}$, final time $T$
        \State $\ket{\psi(T)}\gets  U_\text{eff}(T,0)\,\ket{\psi(0)}$ \Comment{Zero-jump trajectory}
        \State \Call{RecordNorm}{$p_0 = \norm{\psi(T)}^2$}
        \For{$j<N$} \Comment{Number of trajectories}
        \For{$t<T$} \Comment{Operator from jump operators}
            \State Draw a random number $s \in [0, 1]$.
            \State Propagate with $H_\text{eff}$ to time 
            \State $\tau = \arg_\tau\left(\norm{\psi(\tau)}^2 = s\right)$
            \State The jump operator $L_i$ is chosen by calculating individual $\delta_i = \langle \psi(\tau)|L^\dagger_i L_i|\psi(\tau)\rangle$ and drawing a random jump operator index weighted by $\delta_i$.
            \State $\ket{\psi(\tau)} \gets \hat{L}\ket{\psi(\tau)}\big/\sqrt{\bra{\psi(\tau)} \hat{L}^\dagger\hat{L} \ket{\psi(\tau)}}$ \Comment{Jump and normalize}
        \EndFor
        \EndFor
        \State At $t$ of interest, normalize (Eq.~\eqref{equ:rho_n_from_psi})
        \State Calculate density matrix using Eq.~\eqref{equ:SQJ_sum}
    \end{algorithmic}

    \label{algo:QJ}
\end{algorithm}

\section{SQJ} 
\label{app:SQJ}
Here, we elaborate on the SQJ algorithm
In the standard quantum jump method according to Dalibart the instantaneous jump probability in the interval $[t, t+\delta t]$ is given by

\begin{equation}
    \delta p = \delta t \langle L^\dagger L(t)\rangle \quad.
\end{equation}
where $\langle L^\dagger L(t)\rangle = \sum_j\langle L_j^\dagger L_j(t_m)\rangle  $ from which follows the zero jump probability in the interval $[t',t]$ is given by
\begin{equation}
\begin{aligned}
    p^{(0)}_{[t',t]} &= \prod_{t_m}\left(1-\langle L^\dagger L(t)\rangle \delta t\right)\\ 
    &= \exp\left(-\int_{t'}^{t} ds \,\langle L^\dagger L (s)\rangle \right)  
\end{aligned}
\end{equation}
for $\delta t\rightarrow 0$.

The probability density of a jump occurring at time $\tau$ is then given by
\begin{equation}
\begin{aligned}
    \Tilde{p}(\tau) &= \lim_{\delta \tau \rightarrow 0} p^{(0)}_{[0,\tau]}\left(1-p^{(0)}_{[\tau,\tau + \delta \tau]}\right) p^{(0)}_{[\tau,T]}\\
    &=p^{(0)}_{[0,\tau]}\langle L^\dagger L(\tau)\rangle p^{(0)}_{[\tau,T]} \quad .
\end{aligned}
\end{equation}
We have used this probability density DQJ (Eq.~\eqref{equ:ProbabilityDensity}). In first order DQJ, we have assumed there to be a zero probability for a second jump and have therefore neglected the $p^{(0)}_{[\tau,T]} $ term. 

The zero jump probability is numerically given by evolving the initially pure and normalized state with the effective Hamiltonian
\begin{equation}
\begin{aligned}
    \frac{d}{dt}\norm{\psi(t)}^2 &= \bra{\psi(t)}iH_{\text{eff}}^\dagger- iH_{\text{eff}}\ket{\psi(t)}\\
    &= - \bra{\psi(t)} L^\dagger L(t)\ket{\psi(t)}\\
    &= -\langle L^\dagger L(t)\rangle  \norm{\psi}^2\\
    \norm{\psi(t)}^2 &=  \exp\left(-\int_{t'}^{t} ds \,\langle L^\dagger L (s)\rangle \right) \quad,
\end{aligned}
\end{equation}
assuming the initial condition $\norm{\psi(t')}^2 = 1$.

We use the formulation of SQJ within the reduced norm framework for practicality. Here, a random number is drawn from the interval $\epsilon \in [0,1]$ and a jump performed at time $\tau$ once the norm is reduced such that $\epsilon = \norm{\psi(\tau)}^2$. We further adapt it to the weakly dissipative limit to ensure a faithful comparison to our method. Namely, if over the entire integration domain the norm is never reduced below the finite $\norm{\psi(T)}^2$ of the zero-jump trajectory, one can restrict the sampling of $\epsilon$ to $\epsilon \in [\norm{\psi(T)}^2, 1]$ for the first jump of a trajectory. After the first jump has occurred. The density matrix can then be assembled as an equally weighted ensemble of trajectories with at least one jump
\begin{equation}
   \rho(t) = p_0 \rho^{(0)}(t)+(1-p_0) \frac{1}{N}\sum_{j=1}^N|\psi(t)\rangle \langle \psi(t)| \quad .
\label{equ:SQJ_sum}
\end{equation}
The adapted SQJ algorithm is summarized in pseudocode in Algorithm~\ref{algo:QJ}.



\end{document}